\documentstyle[graphicx]{basi}
\input epsf

\newcommand{\phunit}{photons cm$^{-2}$ sr$^{-1}$ s$^{-1}$ \AA$^{-1}$}

\newcommand{\msx} {{\it MSX}}
\newcommand{\micron} {$\mu$m}
\newcommand{\iras} {{\it IRAS}}
\newcommand{\fdg} {\hbox{$.\!\!^\circ$}}
\newcommand{\arcmin} {\hbox{$^\prime$}}
\newcommand{\arcsec} {\hbox{$^{\prime\prime}$}}
\begin{document}

\title [Diffuse UV Emission in Orion]{MSX Observations of Diffuse UV Emission in Orion}
\author[J. Murthy et al]
		{Jayant Murthy,$^1$\thanks{e-mail: jmurthy@yahoo.com}
		  R. C. Henry$^2$, L. J. Paxton$^3$ and S. D. Price$^4$\\
		$^1$Indian Institute for Astronomy,  Koramangala, Bangalore 560 034\\
    		$^2$Department of Physics and Astronomy, The Johns Hopkins University,
		      Baltimore, MD 21218\\
  		$^3$ The Johns Hopkins University Applied Physics Laboratory. Laurel, MD 20723\\
		$^4$  AFRL     Hanscomb AFB, MA 01731\\
       }
\maketitle

   \begin {abstract}
We have observed intense diffuse radiation in the UV (1400 \AA\ 
- 2600 \AA ) from three fields around M42 in Orion. Intensities of 20000 \phunit\
were observed to the east and west of M42 with 8000 \phunit\ south of the nebula.
Enhanced emission, perhaps associated with a nearby complex of molecular clouds
observed in CO,  was detected in one of the fields. The \iras\ 100 \micron\ emission in that
region is highly correlated with the UV intensity with a UV-IR 
ratio of 40 \phunit (MJy sr$^{-1}$)$^{-1}$.
In the other two fields there was no structure in the diffuse emission nor was there any
correlation with the \iras\ emission.
\end {abstract}
\begin {keywords}reflection nebulae; Ultraviolet: ISM - ISM: clouds - dust, extinction
\end {keywords}

%

\section{Introduction}

The {\it Midcourse Space Experiment} (\msx) was a US Ballistic Missile
Defense Organization (BMDO) satellite launched on 26 April 1996, from Vandenberg
AFB into a 900 km altitude, semi-sun synchronous orbit. Mill et al. (1994)
provide an overview of the \msx\ mission and
objectives, which included observations of a variety of natural and
man-made phenomena over a spectral range from the
ultraviolet to the mid-infrared.

A large number of
astronomical observations were made during the mission,
characterizing various components of celestial
backgrounds. Price et al. (1997) describe the objectives of the
11 \msx\ astronomy experiments, designated ``CB" (for Celestial
Background), and also detail the data obtained by SPIRIT III, the infrared
telescope aboard \msx. The ultraviolet and visible sensor suite of
four imagers and five hyperspectral imagers (collectively called
UVISI) on \msx\ are described by Hefferman et al. (1996) and
Paxton et al.  (1996).

In the first part of the mission
(May, 1996 to February, 1997), the infrared instruments had the highest priority and were used 
to scan many interesting regions of the sky. Although the two UV imagers and five
hyperspectral imagers (SPIMs) did take data during this phase, while the IR measurements were
being made, it was only after the cryogen ran out (February 1997) that the full suite
of UVISI sensors were used to begin a systematic survey of
the sky as well as dedicated observations of specific regions. More than 400 observations 
(covering a substantial fraction
of the sky) were obtained by the end of this phase of the mission in December 1997. These experiments
were denoted as CB-10 (a systematic survey of the sky) and CB-11 (observations of specific celestial
targets).

The Orion nebula has long been known to be one of the brightest
regions of diffuse UV emission in the sky since the first (and only) observations 
from a sounding rocket flight by Carruthers and Opal (1977).
Despite considerable internal (instrumental)  scattering, they detected a
diffuse signal amounting to approximately 10\% of the direct
starlight, which they attributed to the scattering of starlight from interstellar
dust in Orion. 

In the present paper, we
describe the three {\em MSX} observations, or ``data collection events'' (referred to as DCEs
hereafter), in which the UVISI sensors were directed toward various
regions in Orion.  We detected intense UV emission in
all three fields. In the following
sections we describe our observations and results.


\section{Observations}

\begin{table}
\centering
 \caption[]{UVISI Instruments}
 \label{uvinst}
 \begin{tabular}{llll}
Instr. & Bandpass$^{\mathrm {a}}$ & Field of View$^{\mathrm {b}}$ & Res.$^{\mathrm {c}}$ \\ \hline
IUW & 1240 - 1380 \AA\ (Filter 3) & $13\fdg1\times 10\fdg3$ & 3\arcmin \\
                            & 1430 -- 1760 \AA\ (Filter 6) &                                                   &     \\
IVW & 4000 - 8500 \AA\ (Filter 3) & $13\fdg1 \times 10\fdg3$ & 3\arcmin \\
IUN    & 2000  - 2500 \AA\  (Filter 3) & $1\fdg5 \times 1\fdg2$ & 20\arcsec \\
                            & 2000 - 2670 \AA\ (Filter 4) &                                                 &       \\
                            & 2230 - 2710 \AA\ (Filter 5) &                                                 &       \\
IVN   & 2900 -- 8270 \AA\                   & $1\fdg5 \times 1\fdg2 $ & 20\arcsec\\
\multicolumn{4}{c}{Spectrographic Imagers (SPIMs)} \\
SPIM1   & 1350 - 1740 \AA\  & $1 ^\circ \times 0\fdg05  $ & 6 \AA \\
SPIM2   & 1640 - 2530 \AA\   & $1^\circ \times 0\fdg1 $ & 7 \AA \\
SPIM3   & 2520 - 3910 \AA\   & $1^\circ \times 0\fdg1 $ & 19 \AA \\
SPIM4   & 3790 - 5850 \AA\   & $1^\circ \times 0\fdg1 $ & 31 \AA \\
SPIM5   & 5810 - 8970 \AA\   & $1^\circ \times 0\fdg1 $ & 40 \AA \\ \hline
\end{tabular}
\begin {list} {}{}
\item[$^{\mathrm {a}}$]The bandpass for the imagers was selectable using a variety of
filters; we have listed only those used in our observations.

\item[$^{\mathrm {b}}$]The field of view for the SPIMs was selectable; only those 
options that were used
in our observations are listed here. In all of our observations the
SPIMs were operated with a spatial scan mirror, resulting in an effective field of view of
$1^\circ \times 1\fdg$ SPIM~1 included a BaF$_2$ filter to block the
atmospheric HI and OI lines.

\item[$^{\mathrm {c}}$]The listed  resolutions are spatial resolutions for the imagers
and spectral resolutions for the SPIMs. In the cryogen phase the
spectral resolution of the SPIMs was half that listed because of
binning forced upon us by data bandwidth restrictions.
\end {list}
\end {table}

\begin{figure}
\centering
\includegraphics[width=9cm]{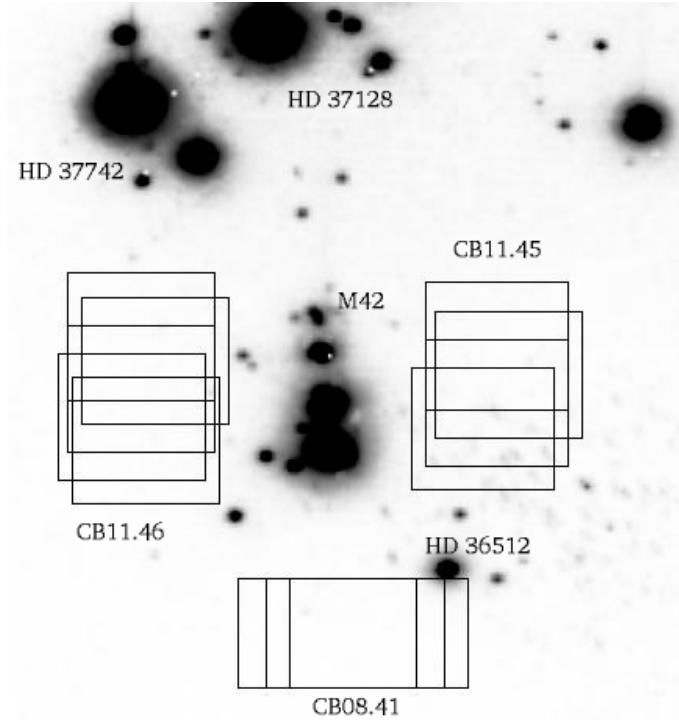}

\caption[]{The positions of the 3 Orion Data Collection Events
(DCES) are shown on an IUW ($13\fdg1\times 10\fdg3$) image (with North
at the top), with a few of the brighter stars
identified: HD 37742 ($\zeta$ Ori) and HD 37128 ($\epsilon$ Ori) form part of the Orion belt.
 There were several pointings within each DCE, with
various filters; the IUN field in each pointing is shown by a small
box ($1\fdg5 \times 1\fdg2$). The SPIMs scanned over essentially the same field as the narrow
field imagers, but with much less sensitivity. Not all of the
instruments and pointings provided useful scientific data (for
our purposes), for instance the IUW image shown was taken at too low a gain
to allow us to detect the diffuse background.}
\label{fig1}
\end{figure}

Because the 9 UVISI instruments were designed to observe
a variety of phenomena, from ballistic missiles in midcourse to the faint diffuse astronomical
radiation, a plethora of filters and operating modes were
available. We list those used in our observations in Table \ref {uvinst}. In the
cryogen phase, when SPIRIT III was operational, the bandwidth
available for the UVISI instruments was limited such that we were only able to download
the data from one imager at a time. However, we could power two (or more) imagers
and download the data from each alternately. In our cryogen phase observations, we routinely
used both the narrow and wide field UV
imagers. with half of the total exposure time in each. The SPIMs took much less
bandwidth and we were able to operate all 5 even in the cryogen phase,
albeit at a reduced spectral and spatial resolution.
There were no bandwidth restrictions in the post-cryogen phase and we operated all
9 instruments at their full resolutions.

All the UVISI instruments used essentially identical detectors ---
intensified CCDs --- with different photocathodes.
Incoming light ejected electrons from the photocathode which
were amplified by a micro-channel plate to strike a green
phosphor coupled by a reducing fiber optic taper to the CCD. The
gain of the MCP was adjustable over a dynamic range of 10$^9$;
however, because we were interested in observations of
relatively faint astronomical sources, we normally operated at
a gain high enough that individual photons could be detected well over the background.
The dynamic range {\it within} a scene was more limited and we could
not always use the photon counting mode because of bright stars
in the field of view. In particular, this was the case for the wide field
imagers in the Orion observations.

The CCDs were read every half second (with an integration time
of 467 milliseconds) and the data were dumped to an on-board tape
recorder. This rapid readout resulted in an enormous amount of
data, typically 1~GB per data collection event (DCE). These data were
processed at the Applied Physics Laboratory of the Johns Hopkins
University under the supervision of the DCATT (Data
Certification and Technology Transfer) team, whose
responsibility it was to certify the process and the data.  The
starting point for our scientific analysis was ``Level 1CA" data:
CCD images with a distortion correction, flat fielding and
calibration already applied.

We observed three regions in Orion, one in the cryogen phase and
the other two after the SPIRIT III instrument was turned off.
These observations are denoted, in temporal order: 

CB08-41 (UJ\_CB08010004101\_05\_9999B),

CB11-45 (UJ\_CB11010004501\_05\_9999A), and
 
CB11-46 (UJ\_CB11010004601\_05\_9999B) 

and are
superimposed on an IUW image of Orion in Fig. \ref{fig1}.  In each DCE,
we observed a specific location for 10 minutes and then offset the pointing by about
$0\fdg25$. This was repeated for a total of four pointings per DCE (three for CB08-41)
with significant field overlap. In each of the two later observations (CB11-45 and CB11-46) filter
changes were made after each pointing. The SPIMs used a scanning mirror to observe a
$1^\circ \times 1^\circ$ field centered on the instrument boresight, overlapping
much of the IUN field of view.

\begin{figure}
\centering
\includegraphics[width=9cm]{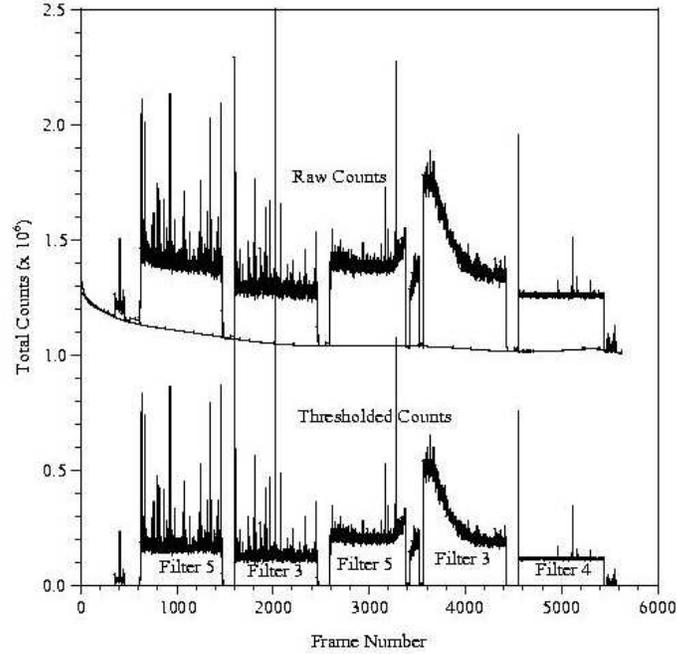}

\caption[]{The count rate for the narrow field of view UV imager (IUN) in CB11-45 as a function 
of frame number is shown. The top line is the raw counts from the imager. From this, we have 
subtracted the CCD read noise (solid line) and thresholded to leave only ``true" photon events
(bottom line). The first 600 frames (5 minutes) were used to measure the dark current 
at different gains after which a series of observations were made using the indicated filters.
The spikes in the data are 
due to bursts of noise in the CCD and were discarded. Note the big rise in the observed counts from 
about frame 3200 to frame 4200, persisting even through a filter closure. Such features are present in 
many of our DCEs and show up in all 9 instruments. We must attribute these to an undetermined 
terrestrial source. The last part of the observation (through Filter 4) was at a lower gain due to the 
presence of a bright star in the field of view. The noise from the photocathode is measured at the 
beginning, middle and end of each observation; in this particular DCE the middle measurement was 
compromised by the general rise in counts.              }
\label{fig2}
\end{figure}

Although the 0.5 second integration time of the UVISI sensors resulted in an enormous amount
of data, the large number of samples in a single DCE were invaluable for diagnostic information.
As an illustrative example, we have plotted (Fig. \ref{fig2}) the total signal per frame from the narrow
field of view imager (IUN) in CB11-45. At the beginning and end of each DCE and at every
filter change the gain on the MCP was set to the minimum value and the observed count rate
dropped to the CCD read noise, setting a floor for the observed signal (square wells in Fig. \ref{fig2}).
The read noise was parametrized and subtracted from the data and the residual was thresholded to leave only
individual photon hits (lower curve in Fig. \ref{fig2}).
A few cosmic ray hits were bright enough to light up the entire detector and were
easily identified (spikes in Fig. \ref{fig2}). In such a case the entire frame was
discarded. Less intense events were also present and were removed by comparing the signal over several
frames.

In addition, there was a general increase in the background level between 
frames 3200 and 4200 in all 9 instruments and persisting even through a filter closure and while
the spacecraft was in an inertial hold. Similar increases in the
total signal have occurred in other DCEs and, in particular,
CB11-46 (executed on the following day) showed a rise at the
same geographical position over central Africa. In other DCEs,
the rise was over other regions of the Earth, typically at low
geographical latitudes and often far from the South Atlantic
Anomaly. Although we have been unable to identify the source of
the signal, it is clearly terrestrial in origin and we
have discarded all affected frames.

\begin{table}
\centering
 \caption[]{Dark Current}
 \label{dk_curr}
 \begin{tabular}{lll}
 & counts pixel$^{-1}$ s$^{-1}$ & Effective Flux$^{\mathrm {a}}$ \\ \hline
CB08-41 (start) & 0.0096 & 1700 \\
CB08-41 (end)  & 0.0034 & 600\\
CB11-45 (start) & 0.025 & 4400\\
CB11-45 (end)  & 0.010 & 1750\\
CB11-46 (start) & 0.025 & 4400\\
CB11-46 (end)  & 0.023 & 4000\\\hline
\end{tabular}
\begin {list} {}{}
\item[$^{\mathrm {a}}$\phunit] This column contains the dark count converted into an effective 
photon flux (using the same calibration as was used for the data).
\end{list}
\end{table}

At this point, we were left with a series of frames containing the astrophysical
signal plus an instrumental dark noise. These frames were added together 
to form a single FITS file for each instrument
and for each pointing. The data product from the imagers was an
image of the region observed and from the SPIMs it was a
3-dimensional file with 2 spatial dimensions and one spectral.

The dark noise in the imagers (at the operational gain) was only read at the beginning
and end of the DCEs (for about a minute each)
and these values are tabulated in Table \ref{dk_curr}. (Note that
in CB11-45 and CB11-46 there actually was another measurement in the middle of the
DCE which was compromised by the general increase in the background level already 
discussed.)  We have used the mean of these two values
as the effective dark current throughout the observation. The deviation
between the two values was added in quadrature to the other errors to yield a total
uncertainty. In all cases the measured dark current was much less than the observed signal.

Because there were far fewer SPIM pixels and because there were no intermediate
points for the CCD read noise parametrization, the background definition in
the SPIMs is very uncertain. Fortunately, 
SPIM 1 included a barium fluoride (BaF$_2$) filter which cut off
all light below 1350 \AA, allowing an absolute measurement of the
effective background level. There was some
wavelength overlap between the different SPIMs and we could
estimate the background subtraction in the other SPIMs by forcing the
spectra to be continuous. However, in practice, because of the
uncertainties in matching the continua and the cumulative errors
in the longer wavelength SPIMs, we could only obtain useful
results for SPIMs 1 and 2, the two UV SPIMs.

\section{Calibration}

\begin{figure}
\centering
\includegraphics[width=9cm]{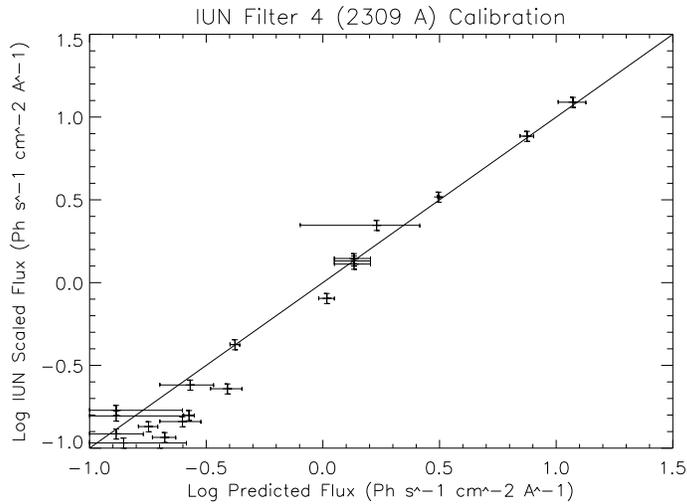}

\caption{The flux observed by IUN filter 4 is compared with predicted fluxes using spectral
types from the SIMBAD database scaled to the flux observed by TD-1. The observed fluxes
match the predicted fluxes over a wide range of brightnesses. We note that these data were taken
from a random sample of fields observed throughout the mission.}
\label{fig3}
\end{figure}
\begin{figure*}

\centering
\includegraphics[width=8cm]{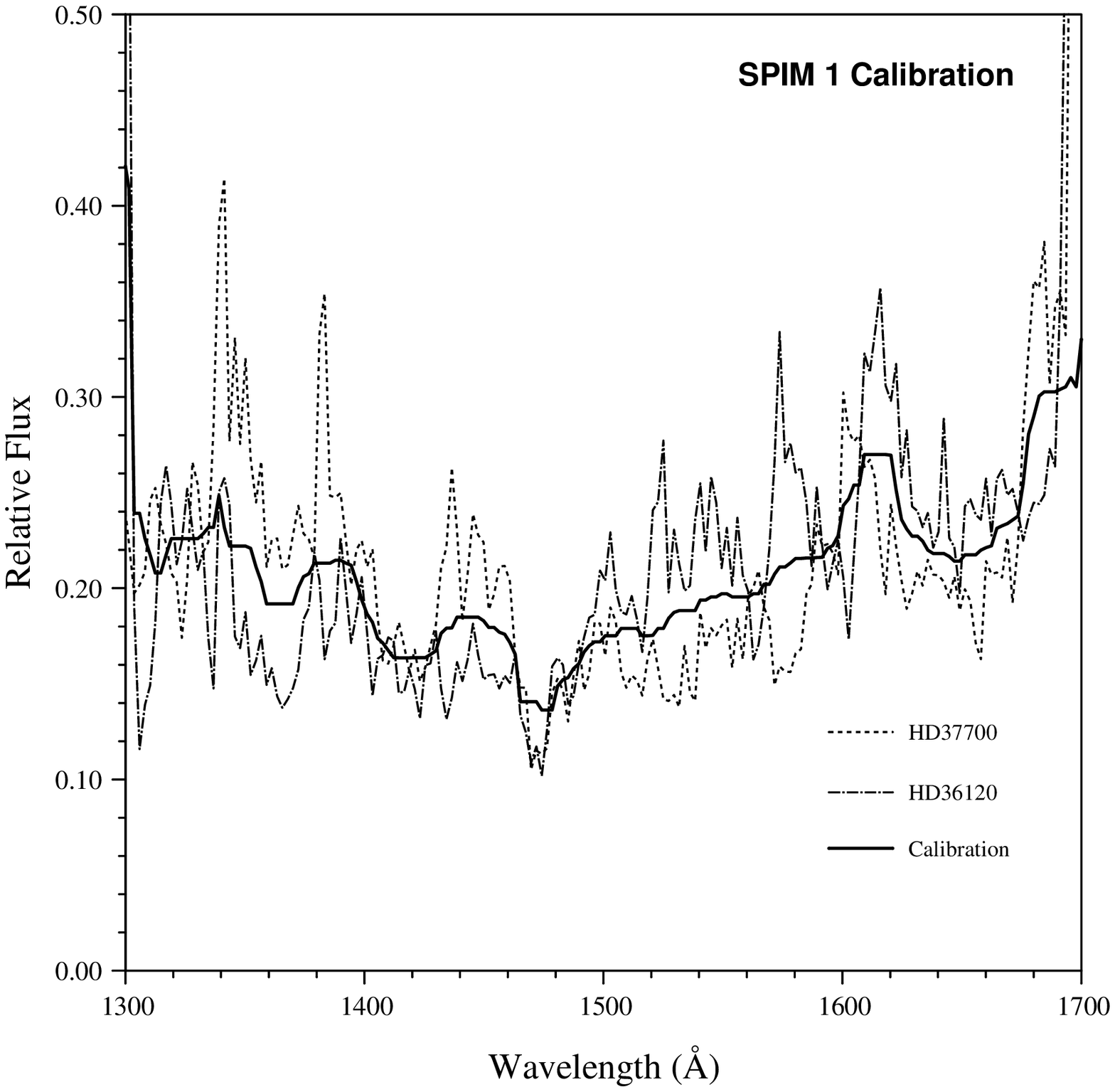}
\includegraphics[width=8cm]{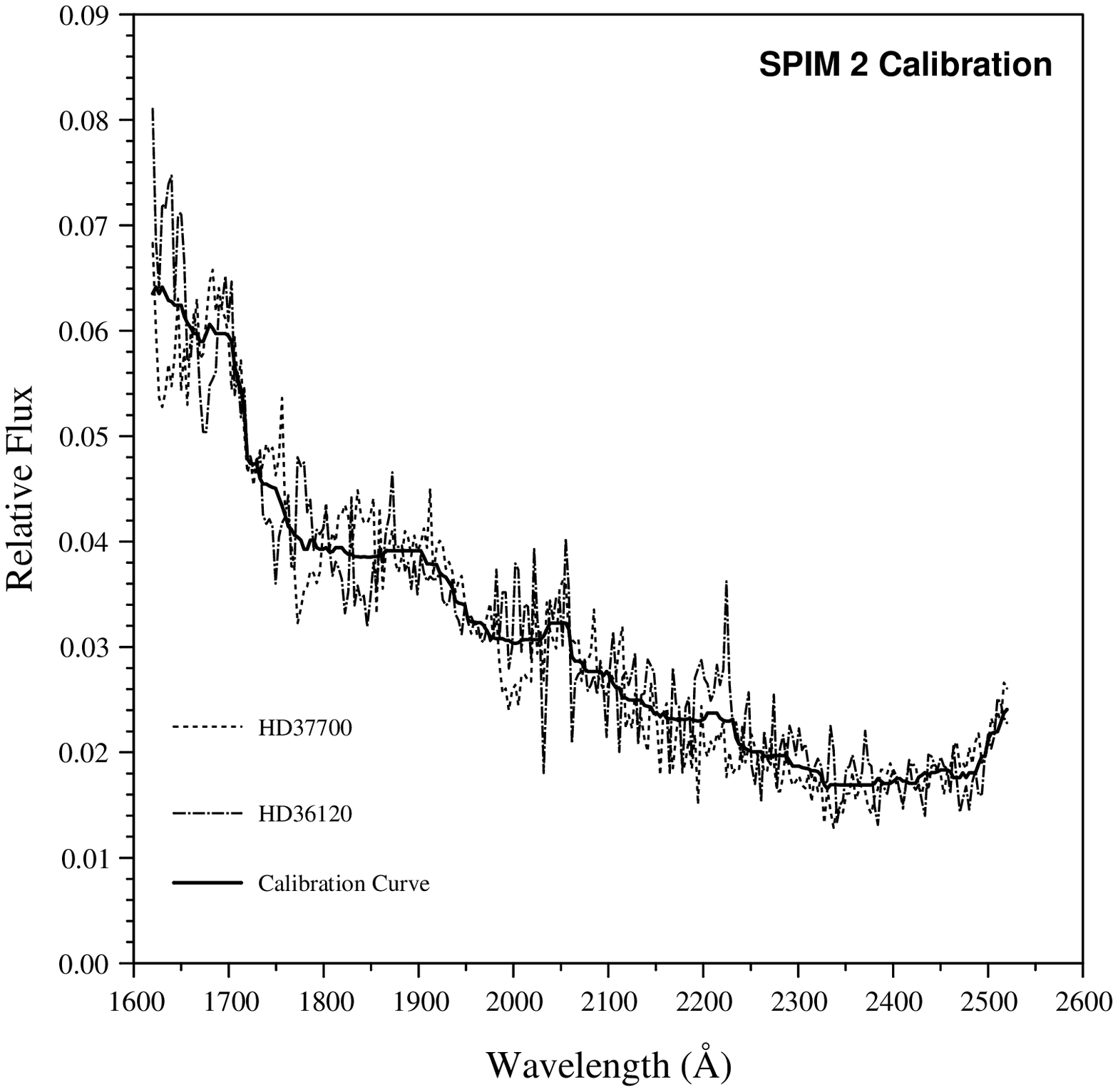}

   \caption{The ratios between the observed SPIM fluxes and the calibrated stellar fluxes as
observed by IUE are plotted for two different stars. Note the generally excellent agreement between the
calibration curves derived from the two stars. We adopted the median filtered average of the two curves as 
our calibration curve. Note that because of background subtraction problems we only calibrated the two shortest 
wavelength SPIMs (SPIMs 1 and 2).}
\label{fig4}
\end{figure*}

The Data Certification and Technology Transfer (DCATT) team had
primary responsibility for characterizing and calibrating the
instrument. However, the calibration in the photon counting mode
was dependent on the details of the processing and we
evolved our own calibration procedures (Newcomer et al. 2001).
We extracted stars from many different DCEs and identified them using the SIMBAD database.
If, and only if, the stars had TD-1 fluxes with reliable spectral types,
we calculated their brightness in each of the instruments and filters and compared with the
observed brightmess. We have plotted one of these calibration curves in Fig. \ref{fig3};
others are shown by Newcomer et al. (2001). In each case, the calibration proved to
be both linear and stable over a large range of brightnesses and
over the lifetime of the satellite. Note that we were forced to reject many of the observed
stars for calibration purposes because of spectral types which were clearly inconsistent
with the TD-1 measurements.

The SPIM calibration is based on two stars (HD 36210 and HD
37700) for which high quality IUE spectra (both SWP and LWP) exist. We
integrated the flux around each star, estimating the background
from the neighbouring pixels, and compared with the archived IUE
spectra. The calibration curves derived from the two stars
(independently) are plotted in Fig. \ref{fig4} and are entirely
consistent with each other. Because our primary spectroscopic science comes from SPIMs~1 and~2,
the two UV SPIMs, 
and because of the background subtraction problems in the longer
wavelength SPIMs, we have focussed on the data from those two spectrographs in this paper.

\section{Results}

\begin{table}
\centering
 \caption[]{Diffuse Radiation in Two Fields}
 \label{background}
 \begin{tabular}{ll}
 & Flux$^{\mathrm {a}}$ \\ \hline
CB08-41 (Filter 4) & $7400 \pm 900 $\\
CB08-41 (Filter 4)  & $7700 \pm 900$\\
CB08-41 (Filter 4)  & $7300 \pm 800$\\
CB11-45 (Filter 3) & $16000 \pm 6000$\\
CB11-45 (Filter 5)  & $20000 \pm 5000$\\\hline
\end{tabular}
\begin {list} {}{}
\item[$^{\mathrm {a}}$\phunit ]
\end{list}
\end{table}

\begin {figure*}
\centering
\includegraphics[width=16cm]{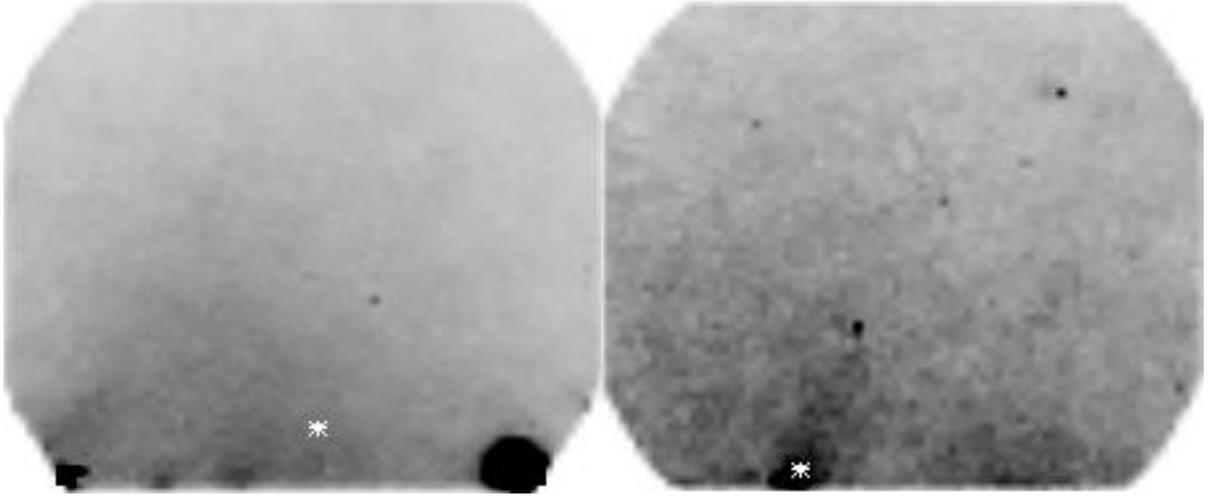}

\caption[]{Images from CB11-46 are shown (North to the right and East to the top)
for IUN Filter 4 (on the left) and IUN Filter 3. The asterisk 
is at the same coordinates in both images  (RA:05 39 29.26 , Dec:-05 41 25.2). A bright patch is visible in 
both images and is probably an extended halo to the complex of molecular clouds identified by Ogura
\& Sugitani (1998) about 
15' to the west (bottom) of the image.}
\label {fig5}
\end {figure*}

\begin {figure}
   \centering
   \includegraphics[width=9cm]{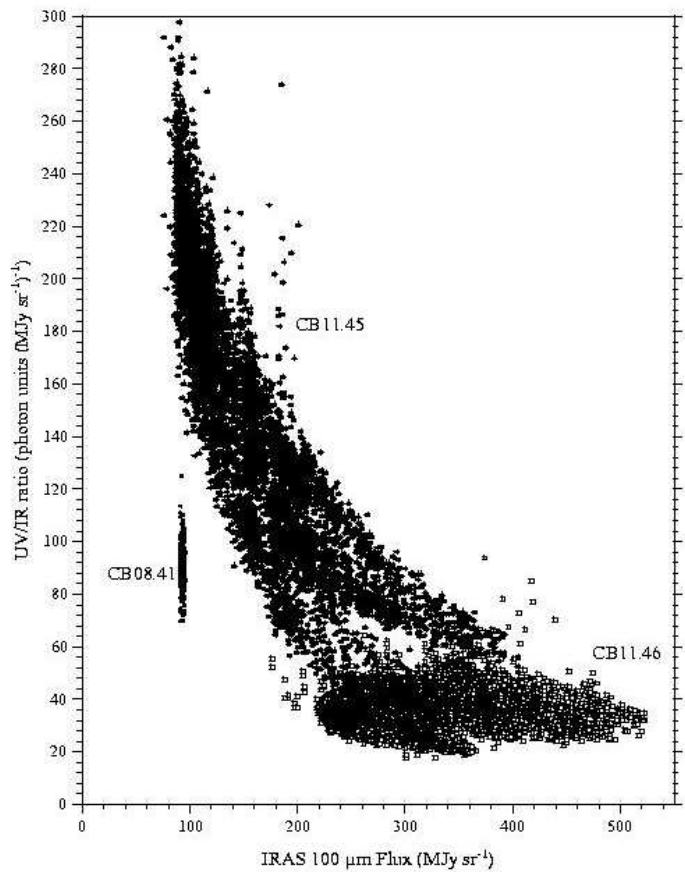}

\caption[]{The correlations between the UV flux observed by IUN (\phunit) and the
IRAS 100 \micron\ flux  MJ sr$^{-1}$) in plotted for the three DCEs observed. In CB08-41 there is not 
much variation in either the IR or the UV fluxes; in CB11-45 the UV flux is almost constant while 
the infrared varies over a factor of 4; and in CB11-46 both vary while keeping the ratio constant.}
\label {fig6}
\end {figure}

\begin {figure}
\centering
\includegraphics[width=9cm]{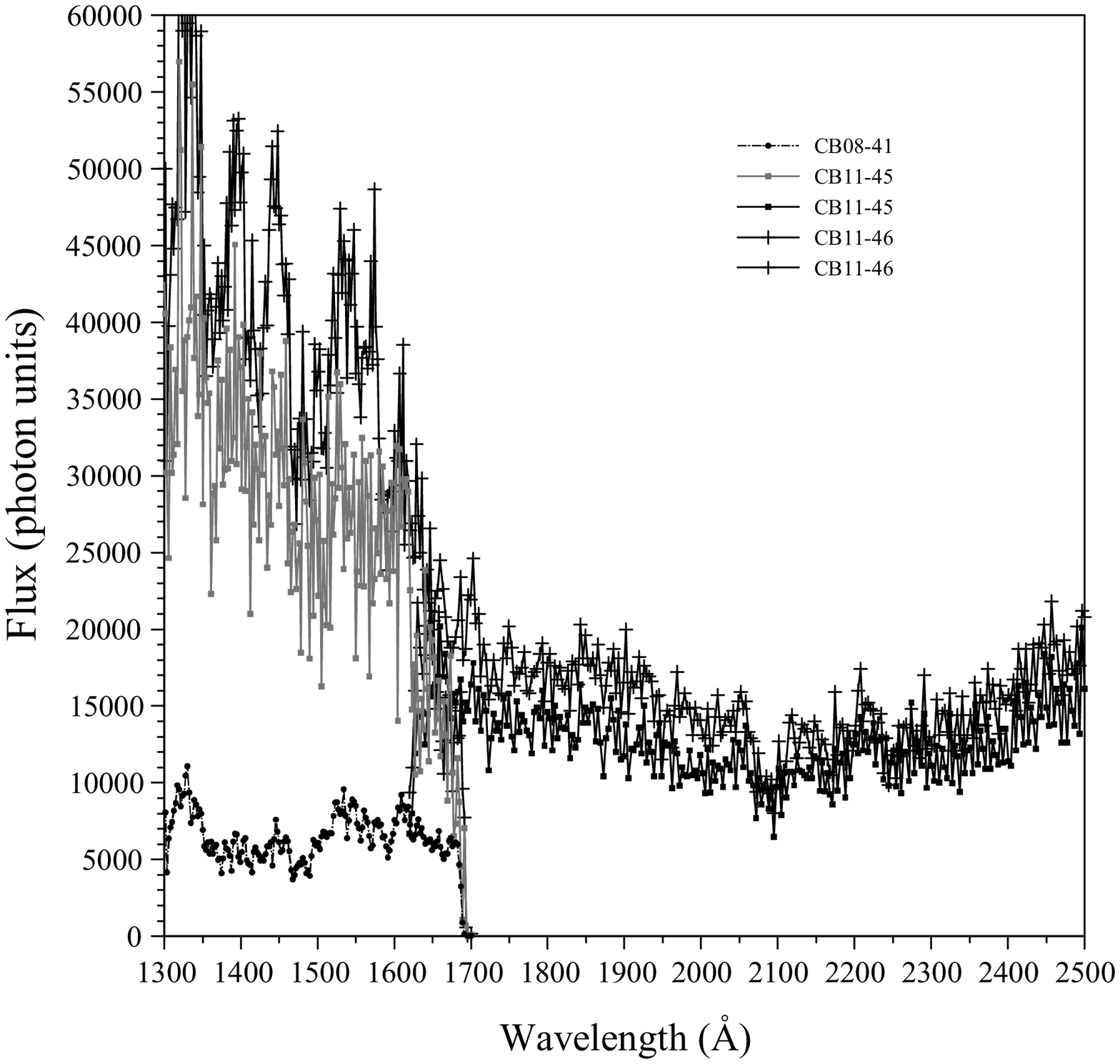}

\caption[]{The spatially integrated spectra observed by the two UV SPIMs in CB11-45 and CB11-46 are
plotted with the legend on the right. Due to processing problems, we could not use the SPIM 2 spectrum 
from CB08-41 and so only the spectrum from SPIM 1 is shown.
Given the different bandpasses, the levels are consistent with the respective IUN fluxes. Note that 
the SPIM coverage in CB11-46 did not include the bright patch observed by IUN}
\label {fig7}
\end {figure}

We observed intense diffuse ultraviolet emission in three regions around M42 in Orion at levels of
more that $2 \times 10^4$  \phunit\ to the east and west of M42. A lower, but still high, level
of 8000  \phunit\ was observed south of the nebula (CB08-41). Instrumental
scattering from the many bright stars in Orion was clearly visible in a few pointings in each DCE
and we have used only those parts of each image which were observed in multiple pointings and in
which the measured flux was consistent between pointings.

In two of the DCEs --- CB08-41 and CB11-45, to the south and west of M42, respectively --- the flux was
uniform with levels tabulated in Table \ref{background}. However, a bright patch was observed in CB11-46 
(Figure \ref{fig5}), most likely emission from the outlying regions of the complex of molecular clouds 
detected by Ogura \& Sugitani (1998)
about 15\arcmin\ away. 

The UV flux is tightly correlated with the 100 \micron\ \iras\ emission
in CB11-46 (Figure \ref{fig6}) with a UV-IR ratio of about 40 \phunit /(MJy sr$^{-1}$)$^{-1}$. The UV to IR ratio from the other two observations is
quite different (also plotted in Fig. \ref{fig6}), particularly for CB11-45 where the
UV flux does not vary at all despite a factor of 4 variation in the
IRAS 100 \micron\ flux.

The only other determination of the UV to IR ratio has come from
Haikala et al (1995)  who used FAUST observations of an isolated
Galactic cirrus cloud to derive a ratio of 128  \phunit
(MJy sr$^{-1}$)$^{-1}$. Although the combination of UV and IR observations of the same
region can provide great insight into the optical properties
of the dust grains, both the UV and the IR emission are highly dependent on the 
exact geometry of the stars and the dust and detailed modeling is required to deconvolve the
properties and nature of the dust.

We also obtained spatial maps of the three DCEs with SPIMs 1 and 2
(due to processing problems only SPIM 1 was available for
CB08-41). There was no evidence for spatial variability in the
diffuse radiation detected by the SPIMs, after discarding the
data near the bright stars in the field, and so we simply
integrated over the entire field of view to give the spectra
plotted in Figure \ref{fig7}. Allowing for the different instrumental
passbands, the diffuse backgrounds detected by the SPIMs are
consistent with those detected by IUN. We note that,
because of a drop in the gain of the SPIMs in the middle of the DCE for CB11-46,
the SPIMs observed only the eastern half of the field.

\section{Conclusions}

We have observed intense diffuse radiation from three fields around
M42 in Orion. This background is much brighter to the east and west of the nebula
with intensities of more that 2 x 10$^4$  \phunit\
dropping to 8000 \phunit\ to the south of M42.
In one of the DCEs (CB11-46 to the east of M42), we detected a bright patch which may
be related to the nearby molecular clouds of Ogura \&
Sugitani (1998). In the other two regions, the emission was uniform.

The UV flux was directly correlated with the IR in CB11-46 with
a ratio of about 40 \phunit\ (MJy sr$^{-1}$)$^{-1}$. No such relation was observed in CB11-45
where, despite a factor of 4 variation in the
\iras\ 100 \micron\ flux, the UV was essentially constant over the field. The only other measurement
of the UV-IR ratio was the 128 \phunit\
(MJy sr$^{-1}$)$^{-1}$ obtained by  Haikala et al (1995) for an isolated
Galactic cirrus cloud. It is clear, and should be
expected, that the UV-IR ratio is heavily dependent on the
environment.

In principle, the combination of IR and UV observations in  a
single direction will strongly constrain the optical properties
of the dust grains. Stellar radiation penetrates into the
interior of the clouds and some part is scattered in the UV.
That part not scattered will be absorbed and then reemitted as
thermal radiation in the IR. In practice, a detailed study is
required to realistically model the penetration and subsequent
reradiation of the input radiation field by the optically thick
(in the UV) dust. We are beginning such a study for simple
geometries where we have Voyager and IRAS data and hope,
eventually, to extend it to more complex regions such as Orion.

This work was supported at the Johns Hopkins University by USAFG F19628-93-K-0004

\end{document}